\documentclass[aps,twocolumn,pra,superscriptaddress,showpacs]{revtex4}
\usepackage{amsmath}
\usepackage{graphicx}
\usepackage{amssymb}
\usepackage{epsfig}
\usepackage{amsfonts}

\begin{document}
\title{Spin Squeezing under Non-Markovian Channels by Hierarchy Equation Method }
\author{Xiaolei Yin}
\affiliation{Advanced Science Institute, RIKEN, Wako-shi, Saitama 351-0198,
Japan}
\affiliation{Zhejiang Institute of Modern Physics, Department of
Physics, Zhejiang University, Hangzhou 310027, China}
\author{Jian Ma}
\affiliation{Advanced Science Institute, RIKEN, Wako-shi, Saitama 351-0198,
Japan}
\affiliation{Zhejiang Institute of Modern Physics, Department of
Physics, Zhejiang University, Hangzhou 310027, China}
\author{Xiaoguang Wang}
\email{xgwang@zimp.zju.edu.cn}
\affiliation{Advanced Science Institute, RIKEN, Wako-shi, Saitama 351-0198,
Japan}
\affiliation{Zhejiang Institute of Modern Physics, Department of
Physics, Zhejiang University, Hangzhou 310027, China}
\author{Franco Nori}
\affiliation{Advanced Science Institute, RIKEN, Wako-shi, Saitama 351-0198,
Japan}
\affiliation{Physics Department, The University of Michigan, Ann Arbor,
MI 48109-1040, USA}
\pacs{03.67. Mn, 03.65. UD, 03.65. Yz}

\begin{abstract}
We study spin squeezing under non-Markovian channels, and consider
an ensemble of $N$ independent spin-$1/2$ particles with exchange
symmetry. Each spin interacts with its own bath, and the baths are
independent and identical. For this kind of open system, the spin
squeezing under decoherence can be investigated from the dynamics of
the local expectations, and the multi-qubit dynamics can be reduced
into the two-qubit one. The reduced dynamics is obtained by the
hierarchy equation method, which is a exact without rotating-wave
and Born-Markov approximation. The numerical results show that the
spin squeezing displays multiple sudden vanishing and revival with
lower bath temperature, and it can also vanish asymptotically.
\end{abstract}

\maketitle

\section{Introduction}

Spin squeezing has attracted much attention for decades~\cite%
{Kitagawa1993,Wineland1994,Sorensen2001,Toth2007,Toth2009,SorensenMolmer,WangSanders,Ma2011}%
. An important application of spin squeezing is to detect quantum
entanglement~\cite{Guehne2009,Amico2008,Horodechi2009}. As a
multipartite entanglement witness, spin squeezing is relatively easy
to be
generated and measured~\cite{Wineland1994,Berman2003,Fernholz2008,Takano2009}%
. Many efforts have been devoted to find relations between spin
squeezing and entanglement~\cite{SorensenMolmer,Sorensen2001,
Kitagawa1993,Wineland1994,Kitagawa2001,WangSanders,Wang2004,Toth2007,Toth2009,Yin2011}%
. Another application of spin squeezing is to improve the precision of
measurements. For example, spin squeezing plays an important role in making
more precise atomic clock~\cite{Wineland1994,
SorensenMolmer,Andre2004,Meiser2008} and gravitational-wave interferometers~%
\cite{Zoller1981, Dunningham,Goda2008}, and so on.

Spin-squeezed states are useful resources for quantum information
processing. However, in practice, decoherence is inevitable and
harmful to spin squeezing and
entanglement~\cite{BreueBook,ZollerBook,
Simon2002,Dur2004,Carvalho2004,Wang2010,Sun2011}. Generally, when
the system-bath coupling strength is weak enough, the decoherence is
studied by using the master equation method, which is derived by
employing the Born approximation~\cite{BreueBook,ZollerBook}.
Besides, the Markov approximation can be applied if the time scale
of the bath is much shorter than that of the system. To overcome the
above approximations, a set of hierarchical
equations were established by Tanimura \textit{et al}~\cite%
{Tanimura1989,Tanimura1990,
Tanimura1991,Tanimura2005,Tanimura2006_1,Tanimura2007,Tanimura2006_2}. It
provides an exact way to obtain the reduced dynamics of system~\cite%
{Tanimura2010,Ma2012}. However, for numerical reasons, it is hard to
treat systems with large number of particles straightforwardly.
Here, we show that for the open system we consider, we can reduce
the multi-particle dynamics to the two-particle one, and then we
efficiently use the hierarchy equation method to make numerical
calculations.

As we know, spin squeezing is a multipartite entanglement witness. Reference~%
\cite{WangMolmer} has shown that for a many-particle system with
exchange symmetry, the spin squeezing parameters of the total system
can be expressed in terms of local expectations and correlations.
Here, we consider such an ensemble of $N$ independent spin-$1/2$
particles. Each particle interacts with its own bath, and the baths
are independent and identical. Thus, the exchange symmetry is not
affected by the decoherence, and the spin squeezing parameters of
the open system can be expressed by dynamics of the local
expectations and correlations. For the system under consideration,
we find that the dynamics of any two particles is governed only by
the local Hamiltonian of the two particles and their baths. Then, we
use the hierarchy equation method to calculate the dynamics the the
local expectations and correlations. Reference~\cite{WangMolmer} has
also shown that the spin squeezing has close relation with pairwise
entanglement if the state of the collective spin system lies in the
$J = N/2$ sector, where $J$ is the collective angular momentum of
the system. Therefore, since the state of the system will not lie in
$J = N/2$ sector anymore under decoherence, the ability of spin
squeezing in detecting pairwise entanglement needs to be further
studied and clarified.

This paper is organized as follows. In Sec.~II, we introduce the
Hamiltonian and the initial state of the open system. The definition
of the spin squeezing parameters is given in Sec.~III, and we also
discuss the symmetry of the system and reduce the multi-qubit
dynamics into the two-qubit one. In Sec.~IV, we introduce the
hierarchy method and give a alternative form of the hierarchy
equation. We numerically calculate spin squeezing parameters and the
rescaled concurrence of the open system under decoherence and
compare their behaviors in Sec.~IV. At last, a summary is given in
Sec.~V.

\section{Hamiltonian and initial state}

The system we consider is an ensemble of $N$ independent spin-$1/2$
particles with exchange symmetry, and each particle interacts with its own
bosonic bath. The $N$ baths are independent and identical. The Hamiltonian
of the total system is ($\hbar$ = 1)
\begin{eqnarray}
H &=&H_{S}+H_{B}+H_{SB}  \notag \\
&=&\sum_{\alpha =1}^{N}\frac{\omega _{0}}{2}\sigma _{\alpha
z}+\sum_{k}\omega _{k}a_{k}^{\dag }a_{k}+  \notag \\
&&\sum_{\alpha=1}^{N}\sum_{k}g_{\alpha k}\sigma _{\alpha x} \left(
a_{k}^{\dag}+a_{k}\right),  \label{hamiltonian}
\end{eqnarray}
where the first term is the Hamiltonian of the system with $\sigma _{k\alpha
} (\alpha = x, y, z)$ the Pauli matrices for the $k$-th spin and $\omega _{0}$ the frequency
for all qubits. The second term describes the bosonic bath, where $b_{k}$
and $b_{k}^{\dagger}$ are the creation and annihilation operators of the $k$%
-th mode with frequency $\omega_{k}$. The system-bath coupling is
characterized by the third term with $g_{\alpha k}$ the coupling
strength for qubit $\alpha$. Here, we study $N$ independent baths,
i.e., the bath can be divided into $N$ parts and $g_{\alpha k}$ is
only non-zero when mode $k$ belongs to the $\alpha $-th part.

The initial state of the total system is set to be a product state
\begin{equation}
\rho _{T}(0)=\rho _{S}(0)\otimes \rho _{B}(0),  \label{rho_tot}
\end{equation}%
where $\rho _{S}(0)$ is a spin-squeezed state and $\rho _{B}(0)$ is a
thermal state given by
\begin{equation}
\rho _{B}(0)=\prod\limits_{k}\frac{\exp (-\beta \omega _{k}a_{k}^{\dag
}a_{k})}{Z_{k}}
\end{equation}%
with the inverse temperature $\beta =1/(k_{B}T)$ and partition function $%
Z_{k}=\mathrm{Tr}\exp (-\beta \omega _{k}a_{k}^{\dag }a_{k})$ for mode $k$.
In this paper we take $k_{B}=1$.

We choose the initial state as a standard one-axis twisted state~\cite%
{Kitagawa1993}
\begin{equation}
|\Psi (0)\rangle =e^{-i\theta J_{x}^{2}/2}| \downarrow ...\downarrow \rangle
\label{one_axis_twis}
\end{equation}
with
\begin{equation}
J_{\alpha }=\frac{1}{2}\sum_{k=1}^{N}\sigma _{k\alpha }
\label{tot_ang_mom}
\end{equation}%
the total angular momentum operators. This state is prepared by the
one-axis twisted Hamiltonian $H=\chi J_{x}^{2}$, with the coupling
constant $\chi$ , and $\theta = 2\chi t$ the twist angle. For our
case, the system of $N$ spin-$1/2$ behaves like an effective large
spin $N/2$.

\section{Spin squeezing and reducing the multi-qubit dynamics into a
two-qubit one}

In this section, we give the definitions of two spin squeezing parameters. By discussing
the symmetry of the open system under consideration, we know that the spin squeezing can be
expressed by the local expectations and correlations. Since we can reduce hte multi-qubit
dynamics into a two-qubit one, the spin squeezing can then calculated by the dynamics of
the local expectations and correlations.

\subsection{Spin squeezing definitions}

There are various measures of spin squeezing related to various
inequality
criteria~\cite{Kitagawa1993,Wineland1994,Sorensen2001,Toth2009,Ma2011},
and we consider two of them as follows:
\begin{eqnarray}
\xi _{\rm KU}^{2} &=&\frac{4(\Delta J_{\perp })_{\min }^{2}}{N},~~ \\
\xi _{\rm T}^{2} &=&\frac{\lambda _{\min }}{\langle \vec{J}^{2}\rangle -\frac{N}{%
2}}.
\end{eqnarray}%
Here, the minimization in the first equation is over all the directions denoted
by $\perp$, which are perpendicular to the mean spin direction $\langle \vec{%
J}\rangle /|\langle \vec{J}\rangle| $. $\lambda _{\min }$ in the second
equation is the minimal eigenvalue of the matrix
\begin{equation}
\Gamma =(N-1)\gamma +\mathbf{C},
\end{equation}%
where
\begin{equation}
\gamma _{kl}=\mathbf{C}_{kl}-\langle J_{k}\rangle \langle J_{l}\rangle,~~~
k,l\in \{x,y,z\}
\end{equation}%
is the covariance matrix and
\begin{equation}
\mathbf{C}_{kl}=\frac{1}{2}\langle J_{l}J_{k}+J_{k}J_{l}\rangle
\end{equation}%
is the global correlation matrix. The parameters $\xi _{\rm KU}^{2}$ and $\xi
_{\rm T}^{2} $ were defined by Kitagawa and Ueda~\cite{Kitagawa1993}, and T\'{o}%
th et al.~\cite{Toth2009}, respectively. If $\xi_{\rm T}^{2}<1$, spin squeezing
occurs, and we can safely say that the multipartite state is entangled~\cite%
{Toth2009,Ma2011}.

From the definitions, we know that the spin squeezing parameters are based on
the expectations and correlations of the collective operators. For the
limitation of the hierarchy equation method, it is hard to calculate the
decoherence of the many-particle system straightforwardly.

\subsection{Simplification of the spin squeezing parameters}
Since the baths are independent and identical, the exchange symmetry
is not affected by decoherence. Therefore, the global expectations
or correlations of collective operators can be written
as~\cite{WangMolmer}
\begin{eqnarray}
\langle J_{\alpha }\rangle &=&\frac{N}{2}\langle \sigma _{1\alpha }\rangle ,
\label{square4} \\
\langle J_{\alpha }^{2}\rangle &=&\frac{N}{4}+\frac{N(N-1)}{4}\langle \sigma
_{1\alpha }\sigma _{2\alpha }\rangle , \\
\langle \lbrack J_{\alpha },J_{\beta }]_{+}\rangle &=&\frac{N(N-1)}{4}%
\langle \lbrack \sigma _{1\alpha },\sigma _{2\beta }]_{+}\rangle,~(\alpha
\neq \beta ),  \label{glob_to_loca}
\end{eqnarray}%
which only depend on the expectation values of the local Pauli operators,
e.g., $\langle \sigma _{1\alpha }\sigma _{2\beta }\rangle$ and $\langle
\sigma _{1\alpha }\rangle$.

The initial one-axis twisted state we use here has a parity symmetry leading
to $\langle J_{x}\rangle =\langle J_{y}\rangle =0,$ namely the mean-spin
direction is along the $z$-axis. Moreover, the mean-spin direction do not change
during decoherence. The proof is given as follows.

The Hamiltonian (\ref{hamiltonian}) displays only one symmetry, i.e., the
parity symmetry. The parity operator is given by
\begin{eqnarray}
\Pi &=&\Pi _{1}\otimes \Pi _{2}  \notag \\
&=&(-1)^{\mathcal{N}}\otimes (-1)^{\sum_{k}a_{k}^{\dagger }a_{k}}  \notag \\
&=&(-1)^{\mathcal{N}+\sum_{k}a_{k}^{\dagger }a_{k}},
\end{eqnarray}%
where $\mathcal{N}=J_{z}+N/2$ describes the numbers of excitations of up
spins. Obviously, we have
\begin{eqnarray}
\Pi H\Pi &=&H, \\
\Pi _{1}\rho _{S}(0)\Pi _{1} &=&\rho _{S}(0), \\
\Pi _{2}\rho _{B}(0)\Pi _{2} &=&\rho _{B}(0), \\
\Pi \rho _{T}(0) \Pi &=&\rho _{T}(0),
\end{eqnarray}%
namely, the Hamiltonian and the initial state have a fixed parity.
Since the exchange symmetry leads to $\langle J_{x}\rangle =N\langle
\sigma _{1x}\rangle/2$, we obtain
\begin{eqnarray}
\langle \sigma _{1x}\rangle &=&\mbox{Tr}\big[\sigma _{1x}U(t)\rho
_{T}(0)U^{\dagger }(t)\big]  \notag \\
&=&\mbox{Tr}\big\{\sigma _{1x}\Pi \big[ \Pi U(t)\Pi \big] \big[ \Pi \rho
_{T}(0)\Pi \big] \big[ \Pi U^{\dagger }(t)\Pi \big] \Pi \big\}  \notag \\
&=&\mbox{Tr}\big[\sigma _{1x}\Pi U(t)\rho _{T}(0)U^{\dagger }(t)\Pi \big]  \notag \\
&=&\mbox{Tr}\big[\Pi \sigma _{1x}\Pi U(t)\rho _{T}(0)U^{\dagger }(t)\big]  \notag \\
&=&-\langle \sigma _{1x}\rangle ,
\end{eqnarray}%
which leads to $\langle J_{x}\rangle =0$. Similarly, $\langle J_{y}\rangle
=\langle J_{y}J_{z}\rangle =\langle J_{x}J_{z}\rangle =0$ can be proved.
Therefore, during the evolution the mean spin direction is always along the $%
z$-axis. In this case, the spin squeezing parameters reduce to~\cite%
{WangSanders,Wang2010}
\begin{eqnarray}
\xi _{\rm KU}^{2} &=& 1+2(N-1)(\langle \sigma _{1+}\sigma _{2-}\rangle -|\langle
\sigma _{1-}\sigma _{2-}\rangle |),  \label{xi1} \\
\xi _{\rm T}^{2} &=&\frac{\min \left \{ \xi _{KU}^{2},\varsigma ^{2}\right \} }{%
(1-1/N)\langle \vec{\sigma}_{1}\cdot \vec{\sigma} _{2}\rangle +1/N},
\label{xi3}
\end{eqnarray}
where
\begin{equation}
\varsigma ^{2} = 1+(N-1)\left( \langle \sigma _{1z}\sigma _{2z}\rangle
-\langle \sigma _{1z}\rangle \langle \sigma _{2z}\rangle \right) .
\label{zita}
\end{equation}

For convenience, hereafter we use
\begin{equation}
\zeta_{k}^{2}=\max (0,1-\xi_{k}^{2}),~~~k\in \{\rm KU,T\}
\end{equation}%
to characterize spin squeezing. With the above definition, spin squeezing
occurs when $\zeta _{k}^{2}>0$.

Now we only need to calculate the dynamics of the local expectations and
correlations of the spins, the spin squeezing parameters are greatly simplified.
However, we still have to use the dynamics of the density matrix of the system
to calculate the local expectations and correlations.

\subsection{Reducing the multi-qubit dynamics into a two-qubit one}

Now we prove that we can reduce the multi-qubit dynamics into a two-qubit one. Generally, we consider a system written as follows
\begin{equation}
H=\sum_{i=1}^{N}H^{(i)},\text{ }H^{(i)}=H_{S}^{(i)}+H_{B}^{(i)}+H_{SB}^{(i)}.
\end{equation}%
$H_{S}^{(i)}$ is the Hamiltonian of one particle, $H_{B}^{(i)}$ is
the bath Hamiltonian, and the couplings are expressed by
$H_{SB}^{(i)}$. Obviously, each of the particles interacts with its
own bath. The particles do not have interaction with each other, and
the baths are independent. Equation~(\ref{hamiltonian}) belongs to
this case.

The time-evolution operator of the total system can be written as
\begin{equation}
U(t)=e^{-iHt}=\prod_{i}e^{-iH_{i}t}=\prod_{i}u_{i}(t),
\end{equation}%
where $u_{i}(t)=e^{-iH_{i}t}$. Then, the total density matrix at time $t$ is
given by
\begin{equation}
\rho _{T}(t)=U(t)\rho _{\mathrm{T}}(0)U^{\dagger }(t),
\end{equation}%
which can be formally written as
\begin{eqnarray}
\rho _{T}(t) &=&U(t)\rho _{T}(0)U^{\dagger }(t)  \notag \\
&=&\prod_{i}u_{i}(t)\rho _{T}(0)\prod_{i}u_{i}^{\dagger }(t).
\end{eqnarray}%
Here we assume that the initial state is a product state written as
\begin{equation}
\rho _{T}(0)=\rho _{S}(0)\otimes \rho _{B}(0).
\label{rho_zero}
\end{equation}%
By tracing out the baths and $N-2$ particles of the system, we obtain the
reduced density matrix of any two particles
\begin{eqnarray}
&&\rho _{S}^{12}(t)  \notag  \label{trac_rho} \\
&=&\mbox{Tr}_{\left\{ B_{1,2}\right\} }\left[ \mbox{Tr}_{\left\{
S_{3...N}B_{3...N}\right\} }\left( \prod_{i=1}^{N}u_{i}(t)\rho _{T%
}(0)\prod_{i=1}^{N}u_{i}^{\dagger }(t)\right) \right]   \notag \\
&=&\mbox{Tr}_{\left\{ B_{1,2}\right\} }\left[ \mbox{Tr}_{\left\{
S_{3...N}B_{3...N}\right\} }\left( \prod_{i=1}^{2}u_{i}(t)\rho _{T%
}(0)\prod_{i=1}^{2}u_{i}^{\dagger }(t)\right) \right]   \notag \\
&=&\mbox{Tr}_{\left\{ B_{1,2}\right\} }\left[ \prod_{i=1}^{2}u_{i}(t)\left(
\rho _{S}^{12}(0)\otimes \rho _{B}^{12}(0)\right)
\prod_{i=1}^{2}u_{i}^{\dagger }(t)\right] ,  \notag \\
&&
\label{rho_redu}
\end{eqnarray}%
where the second equality follows from the fact (see Appendix A)
\begin{eqnarray}
&&\mbox{Tr}_{2}\left[ \left( A_{1}\otimes A_{2}\right) \rho _{12}\left(
B_{1}\otimes B_{2}\right) \right]   \notag \\
&=&\mbox{Tr}_{2}\left[ A_{1}\otimes (B_{2}A_{2})\rho _{12}\left(
B_{1}\otimes I_{2}\right) \right] ,  \label{rho_12}
\end{eqnarray}%
and the last equality is obtained by substituting the initial product state
(\ref{rho_zero}). $\rho _{S}^{12}(0) = {\rm Tr}_{\left\{S_{3...N}\right\}}\rho_{S}(0)$ and $\rho _{B}^{12}(0) = {\rm Tr}_{\left\{B_{3...N}\right\}}\rho_{B}(0)$
in the equation are the reduced density matrices of the initial state for the system and baths respectively.

The above equation (\ref{rho_12}) tells that the evolution of any
two particles is governed only by the local Hamiltonian of the two
particles and their baths. It is noted that we can reach this
conclusion even when the initial state of the system or the baths
are entangled states. Therefore, the multi-qubit dynamics reduces to
the two-qubit one. Then we use the hierarchy equation method to
calculate the two-qubit reduced density matrix of the system, and
the dynamics of the local expectations and correlations in
Eqs.~(\ref{xi1})-(\ref{zita}) can also be obtained.

Here we emphasize that we obtain this conclusion without using exchange symmetry,
which means that the particles are not necessarily identical,
and so do the baths. Also, the proof can be easily extended to any finite number of particles.

\section{Hierarchy equations and initial two-qubit reduced density matrix}

To start with the numerical calculations, we introduce the
hierarchy equation method~\cite{Tanimura2010,Tanimura2006_2}
and discuss the spin squeezing parameters of the initials state in this section.
For comparison, the definition of a rescaled concurrence is also given.

\subsection{Hierarchy equations}

We choose the Drude-Lorentz spectrum,
\begin{equation}
J\left( \omega \right) =\frac{2}{\pi }\frac{\omega \lambda \gamma }{\omega
^{2}+\gamma ^{2}},  \label{Drude_Lorentz_spc}
\end{equation}%
where $\gamma $ represents the width of the spectral distribution of the
bath mode and $\lambda $ can be viewed as the system-bath coupling
strength. The bath correlation function for the bath operator
\begin{equation}
B_{\alpha }(t)=\sum_{k}g_{\alpha k}\left( b_{k}^{\dagger }e^{i\omega
_{k}t}+b_{k}e^{-i\omega _{k}t}\right)
\end{equation}%
is given by~\cite{Tanimura2010}
\begin{equation}
\langle B_{\alpha }(t)B_{\alpha }\left( \tau \right) \rangle
=\sum_{n=0}^{\infty }c_{n}e^{-\nu _{n}|t-\tau |},  \label{eq:Bath_Corr}
\end{equation}%
where
\begin{equation}
\nu _{k}=\frac{2\pi k}{\beta }(1-\delta _{k0})+\gamma \delta _{k0},
\end{equation}%
is the $k$-th Matsubara frequency, and
\begin{equation}
c_{k}=\frac{4\lambda \gamma }{\beta }\frac{\nu _{k}}{\nu _{k}^{2}-\gamma ^{2}%
}(1-\delta _{k0})+\lambda \gamma \left[ \cot \left( \frac{\beta \gamma }{2}%
\right) -i\right] \delta _{k0}
\end{equation}%
are the expansion coefficients.

With the Drude-Lorentz spectrum, the hierarchy equations becomes
\begin{eqnarray}
\dot{\rho}_{\vec{n}} &=&-\left[ iH_{S}^{\times }+(\vec{n}_{1}+\vec{n}%
_{2})\cdot \vec{\nu}\right] \rho _{\vec{n}}  \notag \\
&&-\left( \frac{2\lambda }{\beta \gamma }-i\lambda -\sum_{k=0}^{M}\frac{c_{k}%
}{\nu _{k}}\right) V_{\alpha }^{\times }V_{\alpha }^{\times }\rho _{\vec{n}}
\notag \\
&&-i\sum_{\alpha =1}^{2}\sum_{k=0}^{M}n_{\alpha k}\left( c_{k}V_{\alpha
}\rho _{\vec{n}-\vec{e}_{\alpha k}}-c_{k}^{\ast }\rho _{\vec{n}-\vec{e}%
_{\alpha k}}V_{\alpha }\right)  \notag \\
&&-i\sum_{\alpha =1}^{2}\sum_{k=0}^{M}V_{\alpha }^{\times }\rho _{\vec{n}+%
\vec{e}_{\alpha k}},
\label{hier_equa}
\end{eqnarray}%
where
\begin{equation}
\vec{n}=(\vec{n}_{1},\vec{n}_{2})=(n_{10},...,n_{1M},n_{20},...,n_{2M})
\end{equation}
is a 2$(M+1)$-dimensional vector, a concatenation of two $(M+1)$-dimensional
vectors $\vec{n}_{1}$ and $\vec{n}_{2}.$ The vectors $\vec{\nu}=(\nu
_{0},...\nu _{M})$ and $\vec{e}_{\alpha k}$ are defined as $2(M+1)$%
-dimensional vectors with only $1$ in the $\alpha k$ place and $0$s
in other places. Note that this equation is slightly different and
essentially the same as that given in Ref.~\cite{Tanimura2010}.

\subsection{Initial two-qubit reduced density matrix}

To solve Eq.~(\ref{hier_equa}), we need the initial state. Since the mean spin of the initial state~(\ref{one_axis_twis}) is along the $z$-direction, the two-qubit reduced density matrix can be written
as a block-diagonal form~\cite{WangSanders},
\begin{equation}
\rho _{12}=\left(
\begin{array}{cc}
v_{+} & u^{\ast } \\
u & v_{-}%
\end{array}%
\right) \oplus \left(
\begin{array}{cc}
w & y \\
y & w%
\end{array}%
\right) ,  \label{rho_x}
\end{equation}%
in the basis \{$|00\rangle ,|11\rangle ,|01\rangle ,|10\rangle $\}, where
\begin{eqnarray}
v_{\pm } &=&\left( 1\pm 2\langle \sigma _{1z}\rangle +\langle \sigma
_{1z}\sigma _{2z}\rangle \right) /4,  \label{vpm} \\
w &=&\left( 1-\langle \sigma _{1z}\sigma _{2z}\rangle \right) /4,  \label{r3}
\\
u &=&\langle \sigma _{1-}\sigma _{2-}\rangle ,  \label{rrr} \\
y &=&\langle \sigma _{1+}\sigma _{2-}\rangle .  \label{rr}
\end{eqnarray}%
We notice that if $\langle \sigma _{1+}\sigma _{2-}\rangle ,$ $\langle
\sigma _{1-}\sigma _{2-}\rangle ,$ $\langle \sigma _{1z}\rangle$, and $%
\langle \sigma _{1z}\sigma _{2z}\rangle$ are known, the density matrix is
determined. For the one-axis twisted state, we have~\cite{WangSanders}
\begin{eqnarray}
\langle \sigma _{z}\rangle &=&-\cos ^{N-1}\left( \frac{\theta }{2}\right) ,
\label{mean_sigmaz} \\
\langle \sigma _{1z}\sigma _{2z}\rangle &=&\frac{1}{2}\left( 1+\cos
^{N-2}\theta \right) , \\
\langle \sigma _{1+}\sigma _{2-}\rangle &=&\frac{1}{8}\left( 1-\cos
^{N-2}\theta \right) , \\
\langle \sigma _{1-}\sigma _{2-}\rangle &=&-\frac{1}{8}\left( 1-\cos
^{N-2}\theta \right)  \notag \\
&&-\frac{i}{2}\sin \left( \frac{\theta }{2}\right) \cos ^{N-2}\left( \frac{%
\theta }{2}\right) .  \label{mean_sigmamm}
\end{eqnarray}%
Employing the equations above, we obtain the initial two-qubit reduced
density matrix in Eq.~(\ref{rho_x}). Then we use Eq.~(\ref{hier_equa}) to calculate the dynamics of the reduced density matrix numercially.

Meanwhile, we can also use the Eqs.~(\ref{mean_sigmaz})-(\ref{mean_sigmamm}) to discuss the spin squeezing parameters for the initial state. For the initial state (\ref{one_axis_twis}), we obtain
\begin{eqnarray}
\zeta _{\rm KU}^{2}(0) &=&\zeta _{\rm T}^{2}(0) =\frac{1}{4} \bigg\{\bigg[ (1-\cos
^{N-2}\theta )^{2}+16\sin ^{2}\left( \frac{\theta }{2}\right)  \notag \\
&&\times \cos ^{2N-4}\left( \frac{\theta }{2}\right) \bigg]^{1/2} -1+\cos
^{N-2}\theta \bigg\},  \label{ini_xi}
\end{eqnarray}%
which implies that the two spin squeezing parameters for the initial
state coincide.

It is known that the spin squeezing has close relation with concurrence if the state of the collective spin system lies in the $J = N/2$ sector~\cite{WangMolmer}, such as the initial state of the system. During the decoherence, the state of the system does not lie in $J = N/2$ sector anymore. It is necessary to compare the behaviors of spin squeezing and pairwise entanglement.

The concurrence is defined as~\cite{Wootters}
\begin{equation}
C = \max (0, \lambda_{1}-\lambda_{2}-\lambda_{3}-\lambda_{4}),
\end{equation}
where $\lambda _{1}\geq \lambda _{2}\geq \lambda _{3}\geq \lambda _{4}$ are
the square roots of eigenvalues of $\tilde{\rho}\rho$. Here $\rho$ is the
reduced density matrix of the system, and%
\begin{equation}
\tilde{\rho}=(\sigma_{y}\otimes\sigma_{y})\rho^{\ast}(\sigma_{y}\otimes
\sigma_{y}),  \label{roub}
\end{equation}
where $\rho^{\ast}$ is the conjugate of $\rho$.

For the reduced density matrix of~(\ref{rho_x}), the concurrence is given by~%
\cite{Wootters2000}
\begin{equation}
C = 2\max \left\{ 0,|u|-y,y-\sqrt{v_{+}v_{-}}\right\} .  \label{conc}
\end{equation}
Therefore, we can also obtain the concurrence of the initial state by
employing Eqs.~(\ref{vpm})-(\ref{mean_sigmamm}).

For convenience, here we use a rescaled concurrence
\begin{equation}
C_{r} = (N-1)C,  \label{rsc_ccr}
\end{equation}
and thus $C_{r}(0) = \zeta _{\rm KU}^{2}(0) =\zeta _{\rm T}^{2}(0)$.
Then we know that the two spin squeezing parameters and concurrence
are same for the initial state.

\section{Spin squeezing and concurrence under decoherence}

\begin{figure}[ptb]
\begin{center}
\includegraphics[
width=3.2in]{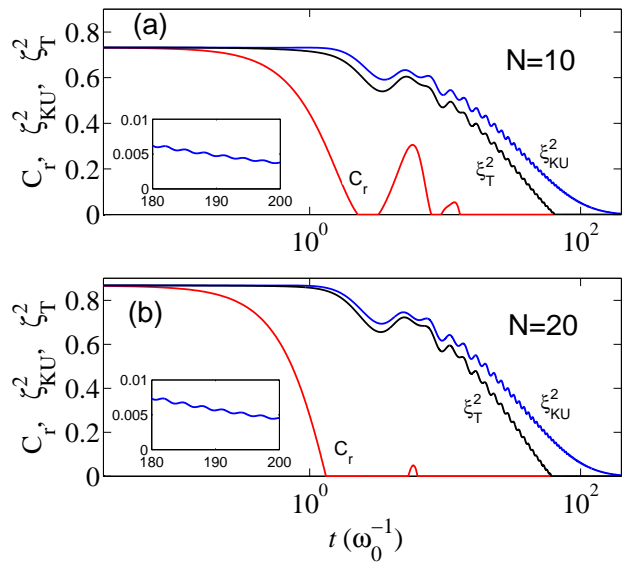}
\end{center}
\caption{(Color online) Time evolution of the spin squeezing parameters $%
\protect\zeta _{\rm KU}^{2}$, $\protect\zeta _{\rm T}^{2}$, and the rescaled
concurrence $C_{r}$ for (a) $N = 10$ and (b) $N = 20$. The inverse
temperature is taken as $\protect\beta = 4/\protect\omega_0$. The insets in
the figures show the magnification of the region where $\protect\zeta %
_{\rm KU}^{2}$ nearly vanishes. The horizontal $x$ axes are logarithmic, but the
inset $x$ axes are linear.}
\label{n10n20}
\end{figure}

The initial one-axis twisted state considered in this work is a symmetric state which can be expressed as a superposition of symmetric Dicke states.
In other words, the $N$ qubits behave effectively like a large spin $N/2$. After
decoherence, not only the symmetric Dicke states will be populated, but also
states with lower symmetry. Therefore, it is not sufficient to describe the system
with only an $(N+1)$-dimensional space. However, the
exchange symmetry is not affected by the decoherence. In other words, a
state with exchange symmetry does not necessarily belong to the
maximally-symmetric space~\cite{Molmer2003}. Now by employing the hierarchy equation method, we calculate the spin squeezing parameters and the rescaled concurrence under decoherence, and compare the behaviors of them.

\begin{figure}[ptb]
\begin{center}
\includegraphics[
width=3.2in ]{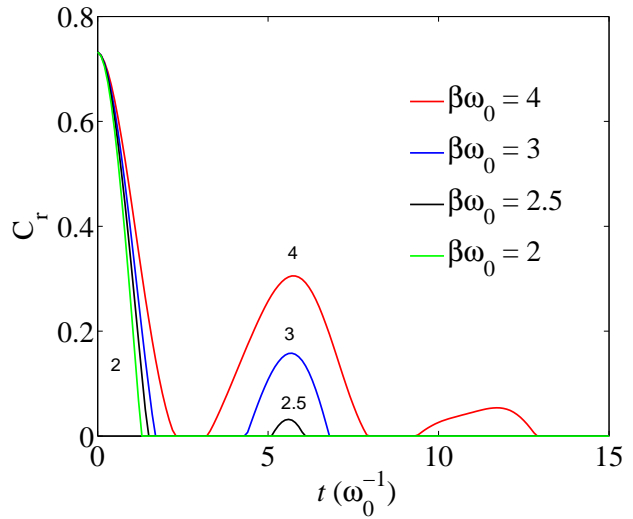}
\end{center}
\caption{(Color online) Time evolution of the rescaled concurrence for
different values of the inverse temperature $\protect\beta$. Here, we choose
$N = 10$.}
\label{conc}
\end{figure}

As an example, we set the initial state given in Eq.~(\ref{ini_xi})
with $\theta = \pi /10$. The parameters of the Drude-Lorentz
spectrum in Eq.~(\ref{Drude_Lorentz_spc}) are chosen to be $\lambda
= 0.03\omega_0$ and $\gamma = 0.15\omega_0$. In this section, we
study the effect of the particle number $N$ and bath temperature $T$
on the dynamics of spin squeezing and concurrence.

Figures~\ref{n10n20}a and~\ref{n10n20}b show the time evolution of $\zeta
_{\rm KU}^{2}$, $\zeta _{\rm T}^{2}$ and $C_{r}$ with two different particle number $N = 10$ and $N = 20$.
The inverse temperature is set to be $\beta = 4/\omega_{0}$. The figures
show that the decay rate of $C_{r}$ increases with $N$. Although the
rescaled concurrence of the initial state for $N = 20$ is larger than that
for $N = 10$, it vanishes earlier. Also, the revival, after a sudden
vanishing, becomes weaker with increasing $N$. Both $%
\zeta _{\rm KU}^{2}$ and $\zeta _{\rm T}^{2}$ decay in an
oscillatory way. We observe that $\zeta _{\rm T}^{2}$ vanishes
suddenly, while interestingly, $\zeta _{\rm KU}^{2}$ decays to zero
asymptotically as shown in the insets. Comparing Fig~\ref{n10n20}a
and~\ref{n10n20}b , we find that for spin squeezing, the vanishing
time changes little with increasing $N$.

\begin{figure}[ptb]
\begin{center}
\includegraphics[
height=2.5217in, width=3.1556in ]{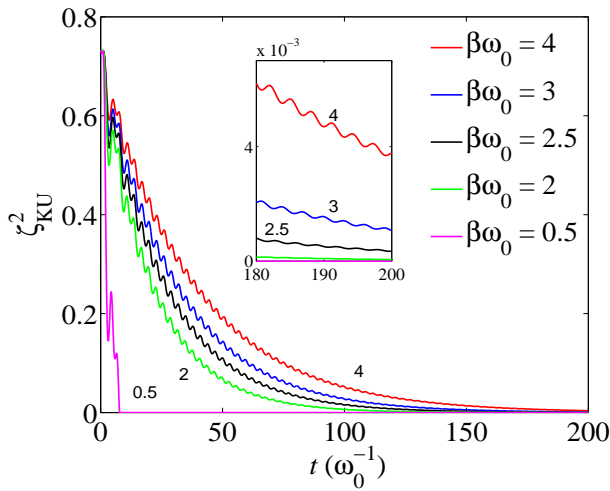}
\end{center}
\caption{(Color online) Time evolution of the spin squeezing parameter $%
\protect\zeta _{\rm KU}^{2}$ for different values of $\protect\beta$, with $N =
10$. The inset shows the magnification of the region when $t$ is large.}
\label{kitagawa}
\end{figure}

Now we focus on the effects of the bath temperature on the dynamics of spin
squeezing and rescaled concurrence, which are shown by Figs.~\ref{conc}-\ref%
{toth}. These figures are plotted with a fixed particle number $N = 10$ and
different temperature $T$. Here we choose the inverse temperature $\beta = 4/\omega_0,
 3/\omega_0, 2.5/\omega_0, 2/\omega_0$, and we specially take $\beta = 0.5/\omega_0$ for
$\zeta_{\rm KU}^{2}$. Firstly, let us discuss the time evolutions of $%
C_{r}$ which are shown in Fig.~\ref{conc}. As expected, $C_{r}$ is
suppressed with increasing temperature. When we choose a low
temperature, such as $\beta = 4/\omega_{0}$, $C_{r}$ decays with
multiple revivals. When the temperature increases, the revivals
become weaker. $C_{r}$ even vanishes completely without revival when
$\beta = 1/\omega_{0}$.

\begin{figure}[ptb]
\begin{center}
\includegraphics[
height=2.5217in, width=3.1556in ]{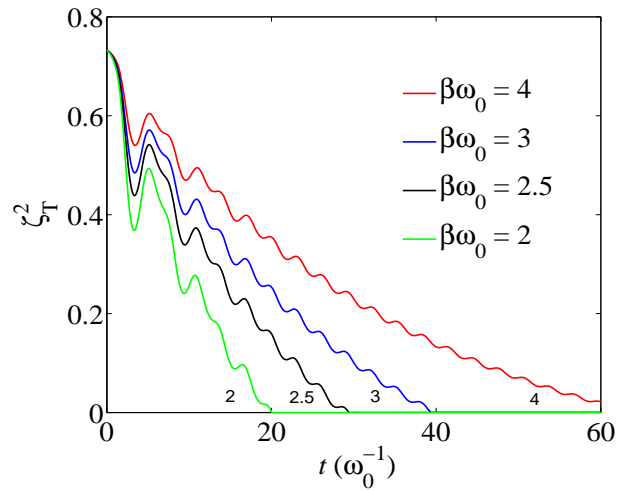}
\end{center}
\caption{(Color online) Time evolution of the spin squeezing parameter $%
\protect\zeta _{\rm T}^{2}$ for different values of $\protect\beta$, with $N = 10
$.}
\label{toth}
\end{figure}

The spin squeezing is also suppressed with increasing $T$. As shown
in Fig.~\ref{kitagawa}, $\zeta_{\rm KU}^{2}$ decays without sudden
vanishing and approaches zero asymptotically ($t \rightarrow
\infty$) when temperature is not high enough, which is shown in the
inset. Interestingly, when temperature reaches to $\beta =
0.5/\omega_0$, $\zeta_{\rm KU}^{2}$ decays to zero quickly and
suddenly without revival. The behavior is quite different with
$C_{r}$. While $\zeta _{\rm T}^{2}$ decays and suddenly vanishes
even with low temperature as shown in Fig.~\ref{toth}, which is
similar to $C_{r}$.

From the comparison, we find that the spin squeezing is not a
satisfactory indicator of pairwise entanglement under decoherence
for the open systemr, although they have close relations. Also, it
is noted that the spin squeezing and concurrence both decay with
oscillations, which is a reflection of the non-Markovian dynamics of
the system.

\section{Conclusion}

In this work, we consider an ensemble of $N$ spin-$1/2$ particles
interacting with identical independent bosonic heat baths. The
one-axis twisted state is chosen to be the initial state. The mean
spin direction of the initial state is along the $z$-axis, and it
does not change during the decoherence dynamics. For the open system
we consider, we proved that the multi-qubit dynamics can be reduced
into a two-qubit one. Then we use the hierarchy equation method to
study the spin squeezing and concurrence under decoherence. This is
an exact method without using rotating-wave and the Born-Markov
approximation.

From the numerical results, we find that the decay rate of the
rescaled concurrence increases with the particle number $N$ as well
as the bath temperature $T$, and the revivals become weaker over
time. For the spin squeezing, it is suppressed with increasing
temperature as expected, while the vanishing time changes little
with $N$. The spin squeezing parameter $\zeta _{\rm KU}^{2}$
vanishes asymptotically with low bath temperature and disappear
suddenly when bath temperature is high enough. Interestingly, $\zeta
_{\rm T}^{2}$ vanishes suddenly even when bath temperature is low,
which is similar to $C_{r}$.

\begin{acknowledgments}
FN acknowledges partial support from the LPS, NSA, ARO, NSF grant No. 0726909,
JSPS-RFBR contract No. 09-02-92114, Grant-in-Aid for Scientific Research (S),
MEXT Kakenhi on Quantum Cybernetics, and the JSPS through its FIRST program.
X. Wang acknowledges support from the NFRPC with Grant No. 2012CB921602 and NSFC
with grant No. 11025527 and 10935010.
\end{acknowledgments}

\appendix

\section{A formula on the partial trace}

It should be noted that the second term of Eq.~(\ref{trac_rho}) is obtained
by moving $\prod_{i=3}^{N}u_{i}^{\dagger }(t)$ from the right of $\rho
_{tot}(0)$ to the left, which use the relation of Eq.~(\ref{rho_12}). We now
prove this property of partial trace. Considering a two-subspace case as an
example:
\begin{eqnarray}
&&\mbox{Tr}_{2}\left[ \left( A_{1}\otimes A_{2}\right) \rho _{12}\left(
B_{1}\otimes B_{2}\right) \right]  \notag \\
&=&\sum_{n}\left[ \left( A_{1}\otimes (\left \langle n\right \vert
A_{2}\right) \rho _{12}\left( B_{1}\otimes (B_{2}\left \vert n\right \rangle
)\right) \right]  \notag \\
&=&\sum_{nm}\left[ A_{1}\otimes (\left \langle n\right \vert A_{2}\rho
_{12}B_{1})\otimes (\left \vert m\right \rangle \left \langle m\right \vert
B_{2}\left \vert n\right \rangle )\right]  \notag \\
&=&\sum_{nm}\left[ \left( A_{1}\otimes (\left \langle m\right \vert
B_{2}\left \vert n\right \rangle \left \langle n\right \vert A_{2}\right)
\rho _{12}B_{1}\otimes \left \vert m\right \rangle \right]  \notag \\
&=&\sum_{m}\left[ A_{1}\otimes (\left \langle m\right \vert B_{2}A_{2})\rho
_{12}B_{1}\otimes \left \vert m\right \rangle \right]  \notag \\
&=&\mbox{Tr}_{2}\left[ A_{1}\otimes (B_{2}A_{2}\rho _{12}\left( B_{1}\otimes
I_{2}\right) \right] ,
\end{eqnarray}%
where $I_{1,2}$ is the identity matrix of the $1$ or $2$ subspace.

\end{document}